\documentclass[twocolumn,showpacs,prd]{revtex4}

\newcommand{\rdot}{{\dot r}}

\usepackage{graphics}

\begin{document}

\title{Numerical simulation of orbiting black holes}

\author{Bernd Br\"ugmann, Wolfgang Tichy, Nina Jansen}

\affiliation{
Center for Gravitational Physics and Geometry and 
Center for Gravitational Wave Physics\\
Penn State University, University Park, PA 16802}

\date{May 12, 2004}

\begin{abstract}
We present numerical simulations of binary black hole systems which
for the first time last for about one orbital period for close but
still separate black holes as indicated by the absence of a common
apparent horizon. An important part of the method is the construction
of comoving coordinates, in which both the angular and radial motion
is minimized through a dynamically adjusted shift condition. We use
fixed mesh refinement for computational efficiency.
\end{abstract}

\pacs{
04.25.Dm, 
04.30.Db, 
95.30.Sf    
%
\quad Preprint number: CGPG-03/12-3
}

\maketitle

One of the fundamental problems of general relativity is the two body
problem of black holes in a binary orbit. Since in general relativity
two orbiting bodies emit gravitational waves that carry away energy
and momentum from the system, the two black holes
spiral inward and eventually merge. Gravitational waves from black
hole mergers are expected to be among the primary sources for
gravitational wave astronomy~\cite{Thorne80,Schutz99}.

The last few orbits of a black hole binary fall into the strongly
dynamic and non-linear regime of general relativity, and we therefore
turn to numerical simulations to solve the full Einstein
equations. Numerical relativity has seen many advances in recent
years, but so far it has not been possible to simulate even a single
binary black hole orbit.
The first 3d simulation of a Schwarzschild black hole was
performed in 1995 \cite{Anninos94c}. In \cite{Bruegmann97}, the first
3d simulation of spinning and moving black holes in a ``grazing
collision'' of near-by black holes inside an innermost stable circular
orbit (ISCO) was presented, see also \cite{Brandt00,Alcubierre00b}.
Simulations starting near or even somewhat
outside an ISCO have been performed e.g.\ in
\cite{Baker:2001nu,Baker00b,Baker:2002qf,Alcubierre2003:pre-ISCO-coalescence-times},
but after rather short evolution times numerical simulations of
black hole binaries become unstable.
In typical advanced simulations the evolution time before
merger is less than $50M$ (where $M$ is the total mass)
\cite{Alcubierre2003:pre-ISCO-coalescence-times}.
An open issue is therefore to find methods that allow longer lasting
evolutions of two black holes before they merge, ideally allowing
evolution times on the order of one or more orbital periods.

In this paper we present results for a new method to choose comoving
coordinates that makes it possible to evolve two black holes for about
one orbital period for the first time. The black holes start out close
to but well outside the ISCO, and the apparent horizons (AHs) do not
merge before one orbital period has passed. Since there are many
different choices for the various components of a numerical relativity
simulation that crucially affect its quality, we will first describe
each one of them in sufficient detail to establish our basic
framework. We will then discuss the major new aspect of our method,
how we construct comoving coordinates, and discuss our numerical
results.

As initial data we choose puncture data \cite{Brandt97b} for two equal
mass black holes without spin on a quasi-circular orbit based on an
approximate helical Killing vector \cite{Tichy03b,Tichy03a}.
Each configuration is determined by the
coordinate distance $\rho_0$ of the punctures from the origin. We
focus on $\rho_0 = 3.0M$, where $M$ is the total ADM
(Arnowitt-Deser-Misner) mass at the punctures. For $\rho_0 = 3.0M$,
the ADM mass at infinity is $0.985M$, the bare mass of one puncture
is $0.477M$, the size of the linear momentum of the individual black
holes is $0.138M$, the angular velocity is $0.0550/M$, and the
orbital period is $T=114M$.  For comparison, post-Newtonian methods
and the thin-sandwich approach find the ISCO in the neighborhood of
$T=65M$ \cite{Damour:2002qh}, which translates to about $\rho_0 =
1.9M$ in our method. The effective
potential method locates the ISCO near $\rho_0=1.1M$, $T=35M$
\cite{Cook94,Baumgarte00a}. 

As evolution system we use the modified version of the
Baumgarte-Shapiro-Shibata-Nakamura (BSSN) system that is described in
detail in \cite{Alcubierre02a}. 
At the outer boundary we impose a radiative boundary condition 
\cite{Alcubierre02a} (we did not implement the monopole term).
The black holes are handled by introducing a time independent excision
boundary according to the ``simple excision'' method described
in~\cite{Alcubierre00a}, with a generalization from
cubical to spherical excision regions. We also perform control
runs without excision using the puncture evolution method
\cite{Bruegmann97,Alcubierre02a}, which typically do not last as long
as the excision runs, but which allow us to check the excision method.

As coordinate conditions we use the dynamic gauge conditions that
proved to be successful for single black hole runs with and without
excision
\cite{Alcubierre00a,Alcubierre01a,Alcubierre02a} and for head-on collisions
\cite{Alcubierre02a}. For the lapse we choose
``1+log'' slicing without explicit shift dependence, and for the shift
we use a particular version of the ``Gamma driver'' condition:
\begin{eqnarray}
\label{gaugea}
\partial_t \alpha &=& - 2 \alpha K \psi^m, \\
\partial_t \beta^i = \frac{3}{4} \alpha^p \psi^{-n} B^i,&& \!\!\!\!\!
\partial_t B^i = \partial_t \tilde\Gamma^i - \eta B^i,
\label{gaugeB}
\end{eqnarray}
where $\alpha$ is the lapse, $\beta^i$ is the shift, $B^i$ is its
first derivative, $K$ is the trace of the extrinsic curvature,
$\tilde\Gamma^i$ is the contracted conformal Christoffel symbol of
BSSN, and $\psi$ is the time independent conformal factor of
Brill-Lindquist data.
After some experimentation we settled for our binary runs on $m=4$,
which helps mimic the singularity avoidance of maximal slicing for
puncture runs, and for the shift we set $n=2$, $p=1$, and $\eta=2/M$.

One important point to be made about the gauge conditions
(\ref{gaugea})-(\ref{gaugeB}) is that although they work well for
black holes without linear momentum, they do not impose corotating or
comoving coordinates. Moving the black hole excision region is
showing a lot of promise \cite{Sperhake:2003fc}, but here we attempt
to minimize the dynamics around black holes at fixed coordinate
positions by modifying the shift condition. 
Corotating frames for numerical relativity are used for example in
\cite{Duez:2002bn} and with dynamic adjustments in
\cite{Alcubierre2003:pre-ISCO-coalescence-times,Alcubierre2003:co-rotating-shift}.
The method that we have developed as a first step toward long term
comoving coordinates is an extension of the methods and ideas of
\cite{Alcubierre2003:pre-ISCO-coalescence-times,Alcubierre2003:co-rotating-shift} to orbiting configurations.

In order to obtain approximately comoving coordinates we introduce the shift
vector 
\begin{equation}
  \beta^i_{com} = \psi^{-q} (A_1 \omega (-y,x,0)^i + A_2 \rdot (-x,-y,0)^i),
\label{com}
\end{equation}
where $x$, $y$, and $z$ denote Cartesian coordinates, with
$\rho = (x^2+y^2)^{1/2}$ and $r=(\rho^2+z^2)^{1/2}$.
The first term inside the brackets is a rotation about the $z$-axis
with angular velocity $A_1 \omega $, while the second term is an inward
radial motion with radial velocity $A_2 \rdot \rho$. The factor $\psi^{-q}$
is used to attenuate the shift to zero at each puncture, which is needed
for simulations without excision. Clearly,
for two point particles on an inspiraling orbit this shift can cancel
the dynamics of the point particles completely. For two orbiting black
holes we can only compensate some aspects of the global motion,
similar to balancing the bulk motion of two stars, with some
dynamics remaining in the metric.

For the runs reported below we have set $q=3$, because this results in
the natural fall off of the shift near punctures \cite{Alcubierre02a},
and we use the same value with excision. 
The prefactor $A_1$ can be used
to attenuate the angular shift for large $r$ 
\cite{Alcubierre2003:pre-ISCO-coalescence-times}, which simplifies the outer
boundary and the analysis at large $r$ at the cost of introducing
additional differential rotation, but for now we work with
$A_1=1$. 
Since $\psi^{-3}$ tends to 1 for $r\rightarrow\infty$, the
shift corresponds to a rigid rotation for large $r$, in particular the
coordinate motion becomes superluminal beyond a light cylinder.
For the radial shift we attenuate with
  $A_2 = (c^2 + 1)^s/[\rho_0(c^2+\rho^2/\rho_0^2)^s]$, which is
constructed such that at the initial radial distance $\rho_0$ to the black
holes the norm of $A_2(x,y,0)^i$ is unity, at the origin the norm is
zero, for large $\rho$ the fall-off is controlled by $s$, and the shape
of the attenuation can be adjusted with $c$. We set 
$c = 1$ and $s = 2$.

To evolve for one orbital time scale it was necessary to introduce a
dynamic control mechanism with time dependent velocities $\omega(t)$
and $\rdot(t)$ in the commotion shift (\ref{com}) (see also
\cite{Alcubierre2003:co-rotating-shift}). In order to estimate changes
in these velocities we define the vector 
$ a^i(t) = \sum (x^i_{puncture} - x^i) \alpha(t) / \sum \alpha(t) , $ 
where the sums run over all points on the excision boundary in the
orbital plane.  The vector $a^i(t)$ points from the center of the
excision region (from the puncture) in the direction into which the
lapse profile has moved off-center.  At finite time intervals $\Delta
t$, we use $a^i(t)$ to compute a velocity correction
\begin{equation}
  \Delta v^i = \left[- \gamma_{damp} \partial_t a^i(t)
                              - k_{drive} a^i(t) \right] \Delta t ,
\label{deltav}
\end{equation}
which is designed to damp out motion in $a^i(t)$ and to drive 
$a^i(t)$ back to zero as in a damped harmonic oscillator. 
In coordinates where the punctures are located on the $y$-axis,
$\Delta v^i$ defines changes in $\omega(t)$ and $\rdot(t)$ by
$\Delta \omega = \Delta v^x / \rho_0$ and
$\Delta \rdot  = \Delta v^y$.
In our case, useful values for the coefficients are $k_{drive}=0.2/M$
and $\gamma_{damp}=5$.

The evolution of the shift proceeds as follows. 
We set the initial lapse to one and initialize the
shift according to (\ref{com}), for example with $\omega = 0.88\Omega$
and $\rdot = 0$ for $\rho_0 = 3M$, where $\Omega$ is the angular
velocity at infinity defined by the initial data. Note that close to
the black holes a correction to $\Omega$ is necessary but not
unexpected. At each time step during the evolution,
we evolve the shift with (\ref{gaugeB}).
First, we evolve for a time interval of $5M$ without any commotion
correction until lapse and shift have gone through
their first rapid evolution to adjust themselves to the presence of
the black holes. After that we compute $\Delta\omega$ and
$\Delta\rdot$ based on (\ref{deltav})
at resolution independent time intervals of $\Delta t = 2M$, which
defines a shift vector $\Delta\beta^i$ according to (\ref{com}).
This shift vector $\Delta\beta^i$ is added to $\beta^i$ everywhere on
the grid, so the shift changes discontinuously at intervals of $\Delta
t$, but we leave the time derivative $B^i$ unchanged.

Assuming a rigidly rotating frame at large distances,
we generalize the radiative boundary condition taking into account that the
scalar wave propagation no longer happens along the radial direction,
and that tensor components have to be rotated to the new frame.
For any tensor $F$ (indices suppressed) the result is
\begin{equation}
\partial_t F =
{\cal L}_\beta F
-  v \frac{x^k}{r} \left(F - F_{\infty}\right)_{,k}  
- v \frac{F - F_{\infty}}{r}, 
\end{equation}
where ${\cal L}$ is the Lie derivative, $v$ is the wave speed, 
and $F_{\infty}$ is the value of $F$ at infinity.
We have experimented with cubical and spherical outer boundaries, where the
latter is expected to have less problems with a global rotation.
A superluminal shift does not
create a problem in our runs with the outer boundary at $24M$, $48M$,
or $96M$, since we can lower the Courant factor in the outer regions
of our fixed mesh refinement grid, which we describe below, by
switching from Berger-Oliger time stepping to uniform time steps.

All evolutions are carried out with a new version of the BAM code
(``bi-functional adaptive mesh'')~\cite{BruegmannInPreparation}, which is
built around an oct-tree, cell-centered adaptive mesh kernel that
currently is functional for fixed mesh refinements (FMR) without
parallelization. Adaptive mesh refinement (AMR) was made famous in
numerical relativity by Choptuik's work on critical collapse
\cite{Choptuik93}, and especially in 3d it can offer enormous savings
over conventional unigrid codes. However, while the basic technical
problem of writing AMR codes has been solved many times, see e.g.\
\cite{AMRweb2} for an overview and
\cite{Diener99,Choi:2003ba,Choptuik:2003ac,Pretorius:2003wc}
for some recent applications in numerical relativity, there have been
only a handful of examples for the full 3d Einstein equations and the
evolution of one~\cite{Bruegmann96,Lanfermann99,Schnetter:2003rb} or
two~\cite{Bruegmann97} black holes. The FMR technique with nested
boxes of~\cite{Bruegmann96} was essential for the
feasibility of the first 3d grazing collision~\cite{Bruegmann97}.

\begin{figure}[t]
\centerline{\resizebox{7cm}{!}{\includegraphics{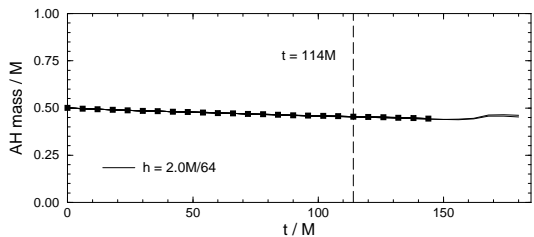}}}
\caption{
Evolution of the AH mass for the black hole binary with
$\rho_0 = 3.0M$. The evolution lasts longer than one orbital period of
$114M$ defined by the initial data. The squares mark a run with $7$
nested levels with coarsest resolution $2M$ and finest resolution 
$h = 0.03125M$, and with the spherical outer boundary at about $48M$,
which crashes around $145M$. Also plotted are results from seven
control runs with the outer boundary at $24M$ and $96M$, with a
cubical outer boundary, and with the AH extracted on a
coarser grid to check its convergence. There is little difference in
the results, except that the runs with the boundary at $24M$ last
somewhat longer.}
\label{fig:ahmass}
\end{figure}

One aspect of the present paper is that we demonstrate that FMR can
work successfully even for black holes in an orbital configuration.
We use nested Cartesian boxes, where for black hole binaries with
equal mass and no spin we only have to store one quadrant of the
global domain.  BAM's Berger-Oliger FMR algorithm uses third order
polynomial interpolation in space and second order polynomial
interpolation in time, following essentially the recipe
of~\cite{Bruegmann96,Bruegmann97}. The main missing feature was a
reasonably stable unigrid code, which is now available in the form of
BSSN with dynamic gauge as discussed above. An important detail of our
setup is the use of the iterative Crank-Nicolson method for time
integration.  To avoid special boundary conditions during
Crank-Nicolson iterations, BAM uses three buffer points
\cite{Schnetter:2003rb}.


\begin{figure}[t]
\resizebox{4cm}{!}{\includegraphics{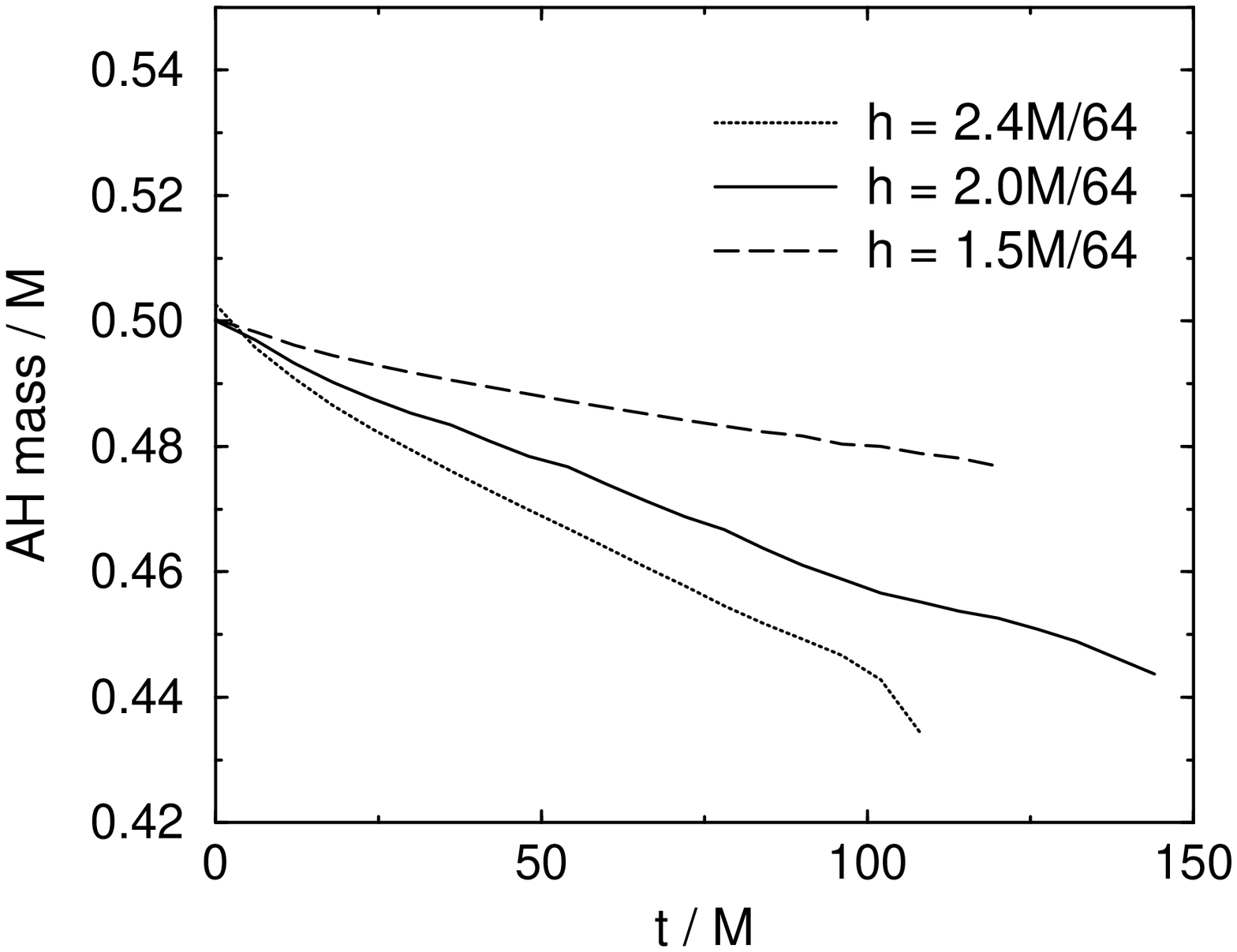}}
\resizebox{3.9cm}{!}{\includegraphics{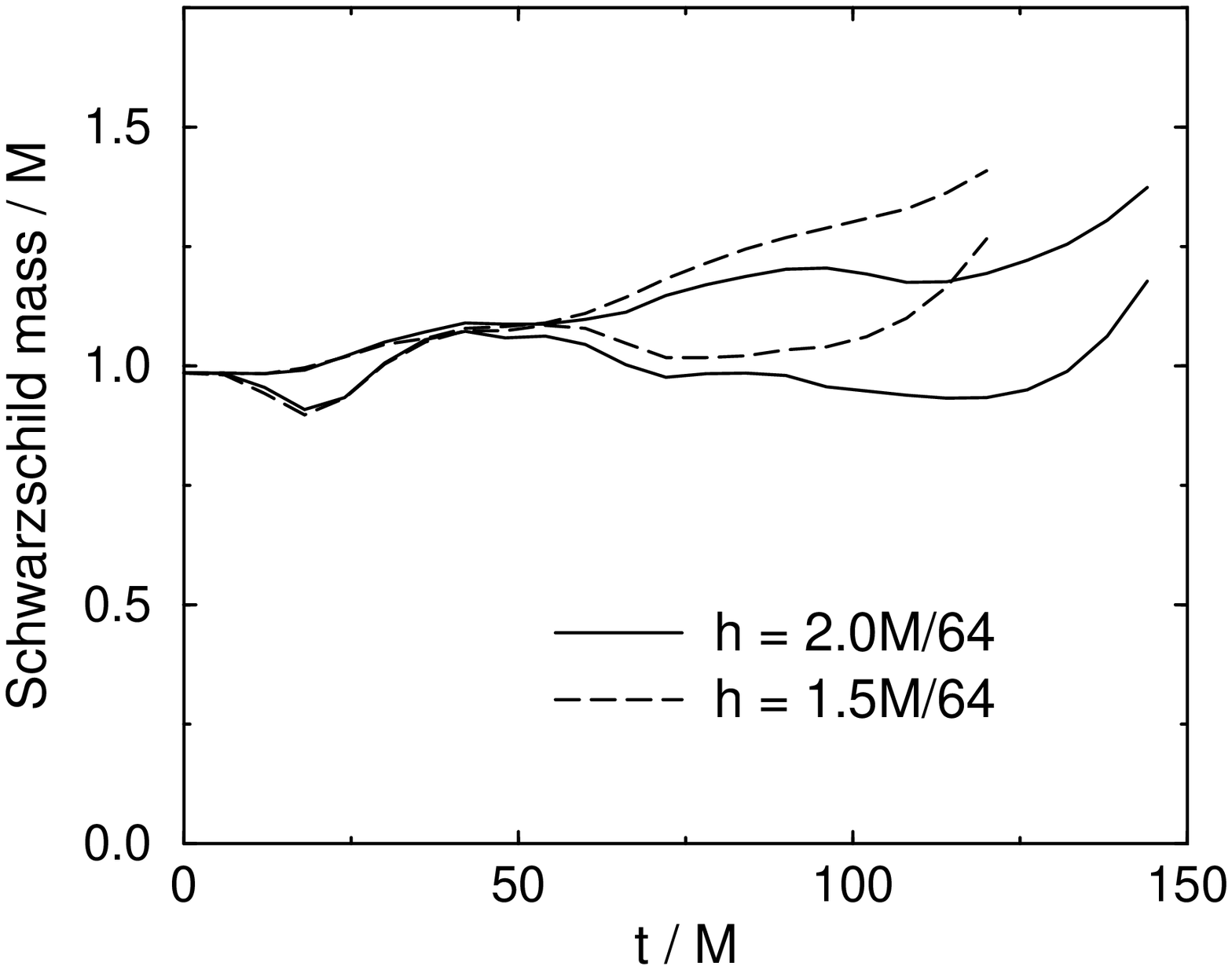}}
\caption{
The panel on the left shows convergence of the AH mass.  The number
and size of the refinement levels was not changed but the overall
resolution was rescaled by a constant factor. There is a linear
downward drift in the mass which becomes smaller with increasing
resolution.  The panel on the right displays the mass at infinity
estimated on a sphere of radius $20M$ assuming a Schwarzschild
background, showing fluctuations of about 20\% to 40\%. The lower and
upper lines for a given resolution correspond to a cubical outer
boundary at $24M$ and $48M$, respectively.  }
\label{fig:convergence}
\end{figure}

\begin{figure}[t]
\vspace{2mm}
\resizebox{7.4cm}{!}{\includegraphics{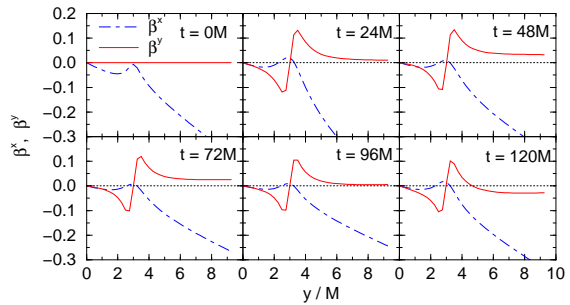}}
\vspace{-2mm}
\caption{
Evolution of the $x$- and $y$-components of the shift vector along the 
$y$-axis. The punctures are located on the $y$-axis at $y=\pm3M$.
}
\label{fig:beta}
\end{figure}

\begin{figure}[t]
\vspace{2mm}
\resizebox{7.2cm}{!}{\includegraphics{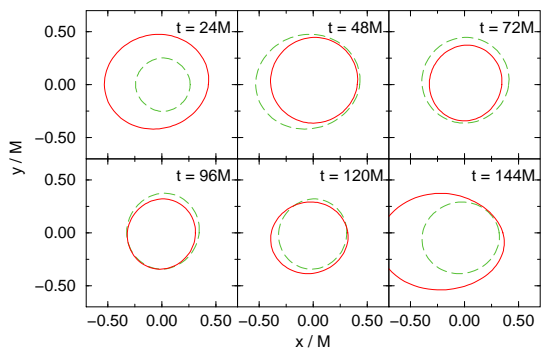}}
\vspace{-2mm}
\caption{
Evolution of the AH of one of the black holes in the $x$-$y$-plane.
The dashed line shows the AH at $t-24M$ in each panel. 
Initially, the AH moves outward quickly while the gauge adjusts itself
near the black hole. It then slowly shrinks toward the center while
being deformed slightly until eventually it drifts out of shape
before the run fails around $145M$. Note that the proper area changes
linearly and only on the order of 10\% during the entire run, see
Fig.~\ref{fig:ahmass}.  }
\label{fig:ahxy}
\end{figure}

Let us summarize our numerical results.  For the black hole binary
with $\rho_0=3.0M$ introduced above, evolution times of up to $185M$
are obtained and a typical run easily exceeds the orbital period of
$114M$. Fig.~\ref{fig:ahmass} shows the AH mass for one
of the black holes as a function of time. It is important to note that
a common AH enclosing both black holes does not form within
the achieved evolution time, while for sufficiently small values of
$\rho_0$ (and the same AH finder described in
\cite{Alcubierre98b} and implemented in Cactus \cite{Cactusweb}) a
common AH is found in
\cite{Alcubierre2003:pre-ISCO-coalescence-times}.

There is an almost linear drift in the AH mass of about 10\% per
$100M$ at a resolution of $h = M/32$ near the excision region, which
becomes smaller with increasing resolution as shown in
Fig.~\ref{fig:convergence}. (We have also evolved Schwarzschild on
quadrants and full grids for $1000M$ and more, confirming that our FMR
method is convergent in the AH mass.)  Puncture evolutions without
excision give a quantitatively very similar result, hence the simple
excision technique does not appear responsible for the drift.  Since
the AH is a slice dependent quantity, the warpage of the slice
contributes to changes in the AH mass. The proper spatial distance
between the AHs along the $y$-axis starts at about $9M$,
rises to $11M$, and drops to $7M$ at $t = 140M$, but since this
distance depends on the gauge and since it does not converge for the
current resolutions, this is only a preliminary result. In the future
we plan to find event horizons to resolve some of the ambiguity.
Fig.~\ref{fig:convergence} also shows an estimate for the mass at
infinity.
The errors are satisfactory for the present purpose. Since
the AH mass shown in Fig.~\ref{fig:ahmass} is not significantly
affected by the location of the outer boundary, we conclude that the
interior of the numerical domain has been computed with good accuracy.

Fig.~\ref{fig:beta} shows the evolution of the shift vector. In
particular, the corotation speed initially increases, then decreases
slowly before increasing again towards the end.  
As an indication of the remaining coordinate motion near the black
holes we show the evolution of the AH in Fig.~\ref{fig:ahxy}. A
residual drift of similar magnitude is observed also for larger
separations, which is a likely reason for the code failure that occurs
after about $150M$ rather independently of separation up to $\rho_0 = 12M$.

In conclusion, dynamically adjusted comoving coordinates enable us to
perform the first numerical simulations of two black holes near but
outside the ISCO for about one orbital period.
A good indicator for one orbit would be the presence of two cycles of
gravitational waves. First experiments with wave extraction indicate
that improvements of the outer boundary are needed.

It is a pleasure to thank M. Alcubierre, P. Diener, N. Dorband, S. Hawley,
D. Pollney, E. Seidel, and the AEI team for many discussions and
collaborations on the methods on which our simulations are based.
S. Hawley performed first experiments with a dynamic corotation shift 
in the context of \cite{Alcubierre2003:pre-ISCO-coalescence-times}. 
We also like to thank A. Ashtekar and P. Laguna for discussions, and
we thank B. Lacki for help with HDF5. We acknowledge the support of
the Center for Gravitational Wave Physics funded by the National
Science Foundation under Cooperative Agreement PHY-01-14375. This work
was also supported by NSF grants PHY-02-18750 and PHY-02-44788.

\bibliography{references}

\begin{thebibliography}{34}
\expandafter\ifx\csname natexlab\endcsname\relax\def\natexlab#1{#1}\fi
\expandafter\ifx\csname bibnamefont\endcsname\relax
  \def\bibnamefont#1{#1}\fi
\expandafter\ifx\csname bibfnamefont\endcsname\relax
  \def\bibfnamefont#1{#1}\fi
\expandafter\ifx\csname citenamefont\endcsname\relax
  \def\citenamefont#1{#1}\fi
\expandafter\ifx\csname url\endcsname\relax
  \def\url#1{\texttt{#1}}\fi
\expandafter\ifx\csname urlprefix\endcsname\relax\def\urlprefix{URL }\fi
\providecommand{\bibinfo}[2]{#2}
\providecommand{\eprint}[2][]{\url{#2}}

\bibitem[{\citenamefont{Thorne}(1980)}]{Thorne80}
\bibinfo{author}{\bibfnamefont{K.}~\bibnamefont{Thorne}},
  \bibinfo{journal}{Rev. Mod. Phys.} \textbf{\bibinfo{volume}{52}},
  \bibinfo{pages}{285} (\bibinfo{year}{1980}).

\bibitem[{\citenamefont{Schutz}(1999)}]{Schutz99}
\bibinfo{author}{\bibfnamefont{B.}~\bibnamefont{Schutz}},
  \bibinfo{journal}{Class. Quantum Grav.} \textbf{\bibinfo{volume}{16}},
  \bibinfo{pages}{A131} (\bibinfo{year}{1999}).

\bibitem[{\citenamefont{Anninos et~al.}(1995)\citenamefont{Anninos, Camarda,
  Mass{\'o}, Seidel, Suen, and Towns}}]{Anninos94c}
\bibinfo{author}{\bibfnamefont{P.}~\bibnamefont{Anninos}},
  \bibinfo{author}{\bibfnamefont{K.}~\bibnamefont{Camarda}},
  \bibinfo{author}{\bibfnamefont{J.}~\bibnamefont{Mass{\'o}}},
  \bibinfo{author}{\bibfnamefont{E.}~\bibnamefont{Seidel}},
  \bibinfo{author}{\bibfnamefont{W.-M.} \bibnamefont{Suen}}, \bibnamefont{and}
  \bibinfo{author}{\bibfnamefont{J.}~\bibnamefont{Towns}},
  \bibinfo{journal}{Phys. Rev. D} \textbf{\bibinfo{volume}{52}},
  \bibinfo{pages}{2059} (\bibinfo{year}{1995}).

\bibitem[{\citenamefont{Br{\"u}gmann}(1999)}]{Bruegmann97}
\bibinfo{author}{\bibfnamefont{B.}~\bibnamefont{Br{\"u}gmann}},
  \bibinfo{journal}{Int. J. Mod. Phys. D} \textbf{\bibinfo{volume}{8}},
  \bibinfo{pages}{85} (\bibinfo{year}{1999}).

\bibitem[{\citenamefont{Brandt et~al.}(2000)\citenamefont{Brandt, Correll,
  G\'omez, Huq, Laguna, Lehner, Marronetti, Matzner, Neilsen, Pullin
  et~al.}}]{Brandt00}
\bibinfo{author}{\bibfnamefont{S.}~\bibnamefont{Brandt}},
  \bibinfo{author}{\bibfnamefont{R.}~\bibnamefont{Correll}},
  \bibinfo{author}{\bibfnamefont{R.}~\bibnamefont{G\'omez}},
  \bibinfo{author}{\bibfnamefont{M.~F.} \bibnamefont{Huq}},
  \bibinfo{author}{\bibfnamefont{P.}~\bibnamefont{Laguna}},
  \bibinfo{author}{\bibfnamefont{L.}~\bibnamefont{Lehner}},
  \bibinfo{author}{\bibfnamefont{P.}~\bibnamefont{Marronetti}},
  \bibinfo{author}{\bibfnamefont{R.~A.} \bibnamefont{Matzner}},
  \bibinfo{author}{\bibfnamefont{D.}~\bibnamefont{Neilsen}},
  \bibinfo{author}{\bibfnamefont{J.}~\bibnamefont{Pullin}},
  \bibnamefont{et~al.}, \bibinfo{journal}{Phys. Rev. Lett.}
  \textbf{\bibinfo{volume}{85}}, \bibinfo{pages}{5496} (\bibinfo{year}{2000}).

\bibitem[{\citenamefont{Alcubierre
  et~al.}(2001{\natexlab{a}})\citenamefont{Alcubierre, Benger, Br\"ugmann,
  Lanfermann, Nerger, Seidel, and Takahashi}}]{Alcubierre00b}
\bibinfo{author}{\bibfnamefont{M.}~\bibnamefont{Alcubierre}},
  \bibinfo{author}{\bibfnamefont{W.}~\bibnamefont{Benger}},
  \bibinfo{author}{\bibfnamefont{B.}~\bibnamefont{Br\"ugmann}},
  \bibinfo{author}{\bibfnamefont{G.}~\bibnamefont{Lanfermann}},
  \bibinfo{author}{\bibfnamefont{L.}~\bibnamefont{Nerger}},
  \bibinfo{author}{\bibfnamefont{E.}~\bibnamefont{Seidel}}, \bibnamefont{and}
  \bibinfo{author}{\bibfnamefont{R.}~\bibnamefont{Takahashi}},
  \bibinfo{journal}{Phys. Rev. Lett.} \textbf{\bibinfo{volume}{87}},
  \bibinfo{pages}{271103} (\bibinfo{year}{2001}{\natexlab{a}}),
  \eprint{gr-qc/0012079}.

\bibitem[{\citenamefont{Baker et~al.}(2001)\citenamefont{Baker, Br{\"u}gmann,
  Campanelli, Lousto, and Takahashi}}]{Baker:2001nu}
\bibinfo{author}{\bibfnamefont{J.}~\bibnamefont{Baker}},
  \bibinfo{author}{\bibfnamefont{B.}~\bibnamefont{Br{\"u}gmann}},
  \bibinfo{author}{\bibfnamefont{M.}~\bibnamefont{Campanelli}},
  \bibinfo{author}{\bibfnamefont{C.~O.} \bibnamefont{Lousto}},
  \bibnamefont{and}
  \bibinfo{author}{\bibfnamefont{R.}~\bibnamefont{Takahashi}},
  \bibinfo{journal}{Phys. Rev. Lett.} \textbf{\bibinfo{volume}{87}},
  \bibinfo{pages}{121103} (\bibinfo{year}{2001}),
  \eprint[http://arXiv.org/abs]{gr-qc/0102037}.

\bibitem[{\citenamefont{Baker et~al.}(2000)\citenamefont{Baker, Br\"ugmann,
  Campanelli, and Lousto}}]{Baker00b}
\bibinfo{author}{\bibfnamefont{J.}~\bibnamefont{Baker}},
  \bibinfo{author}{\bibfnamefont{B.}~\bibnamefont{Br\"ugmann}},
  \bibinfo{author}{\bibfnamefont{M.}~\bibnamefont{Campanelli}},
  \bibnamefont{and} \bibinfo{author}{\bibfnamefont{C.~O.}
  \bibnamefont{Lousto}}, \bibinfo{journal}{Class. Quantum Grav.}
  \textbf{\bibinfo{volume}{17}}, \bibinfo{pages}{L149} (\bibinfo{year}{2000}).

\bibitem[{\citenamefont{Baker et~al.}(2002)\citenamefont{Baker, Campanelli,
  Lousto, and Takahashi}}]{Baker:2002qf}
\bibinfo{author}{\bibfnamefont{J.}~\bibnamefont{Baker}},
  \bibinfo{author}{\bibfnamefont{M.}~\bibnamefont{Campanelli}},
  \bibinfo{author}{\bibfnamefont{C.~O.} \bibnamefont{Lousto}},
  \bibnamefont{and}
  \bibinfo{author}{\bibfnamefont{R.}~\bibnamefont{Takahashi}},
  \bibinfo{journal}{Phys. Rev.} \textbf{\bibinfo{volume}{D65}},
  \bibinfo{pages}{124012} (\bibinfo{year}{2002}),
  \eprint[http://arXiv.org/abs]{astro-ph/0202469}.

\bibitem[{\citenamefont{Alcubierre
  et~al.}(2004{\natexlab{a}})\citenamefont{Alcubierre, Br\"ugmann, Diener,
  Guzm\'an, Hawke, Hawley, Herrmann, Koppitz, Pollney, Seidel
  et~al.}}]{Alcubierre2003:pre-ISCO-coalescence-times}
\bibinfo{author}{\bibfnamefont{M.}~\bibnamefont{Alcubierre}},
  \bibinfo{author}{\bibfnamefont{B.}~\bibnamefont{Br\"ugmann}},
  \bibinfo{author}{\bibfnamefont{P.}~\bibnamefont{Diener}},
  \bibinfo{author}{\bibfnamefont{F.~S.} \bibnamefont{Guzm\'an}},
  \bibinfo{author}{\bibfnamefont{I.}~\bibnamefont{Hawke}},
  \bibinfo{author}{\bibfnamefont{S.}~\bibnamefont{Hawley}},
  \bibinfo{author}{\bibfnamefont{F.}~\bibnamefont{Herrmann}},
  \bibinfo{author}{\bibfnamefont{M.}~\bibnamefont{Koppitz}},
  \bibinfo{author}{\bibfnamefont{D.}~\bibnamefont{Pollney}},
  \bibinfo{author}{\bibfnamefont{E.}~\bibnamefont{Seidel}},
  \bibnamefont{et~al.} (\bibinfo{year}{2004}{\natexlab{a}}), \bibinfo{note}{in
  preparation}.

\bibitem[{\citenamefont{Brandt and Br{\"u}gmann}(1997)}]{Brandt97b}
\bibinfo{author}{\bibfnamefont{S.}~\bibnamefont{Brandt}} \bibnamefont{and}
  \bibinfo{author}{\bibfnamefont{B.}~\bibnamefont{Br{\"u}gmann}},
  \bibinfo{journal}{Phys. Rev. Lett.} \textbf{\bibinfo{volume}{78}},
  \bibinfo{pages}{3606} (\bibinfo{year}{1997}).

\bibitem[{\citenamefont{Tichy et~al.}(2003)\citenamefont{Tichy, Br\"ugmann, and
  Laguna}}]{Tichy03a}
\bibinfo{author}{\bibfnamefont{W.}~\bibnamefont{Tichy}},
  \bibinfo{author}{\bibfnamefont{B.}~\bibnamefont{Br\"ugmann}},
  \bibnamefont{and} \bibinfo{author}{\bibfnamefont{P.}~\bibnamefont{Laguna}},
  \bibinfo{journal}{Phys. Rev. D} \textbf{\bibinfo{volume}{68}},
  \bibinfo{pages}{064008} (\bibinfo{year}{2003}),
  \bibinfo{note}{gr-qc/0306020}.

\bibitem[{\citenamefont{Tichy and Br{\"u}gmann}(2003)}]{Tichy03b}
\bibinfo{author}{\bibfnamefont{W.}~\bibnamefont{Tichy}} \bibnamefont{and}
  \bibinfo{author}{\bibfnamefont{B.}~\bibnamefont{Br{\"u}gmann}}
  (\bibinfo{year}{2003}), \bibinfo{note}{gr-qc/0307027}.

\bibitem[{\citenamefont{Damour et~al.}(2002)\citenamefont{Damour, Gourgoulhon,
  and Grandclement}}]{Damour:2002qh}
\bibinfo{author}{\bibfnamefont{T.}~\bibnamefont{Damour}},
  \bibinfo{author}{\bibfnamefont{E.}~\bibnamefont{Gourgoulhon}},
  \bibnamefont{and}
  \bibinfo{author}{\bibfnamefont{P.}~\bibnamefont{Grandclement}},
  \bibinfo{journal}{Phys. Rev.} \textbf{\bibinfo{volume}{D66}},
  \bibinfo{pages}{024007} (\bibinfo{year}{2002}),
  \eprint[http://arXiv.org/abs]{gr-qc/0204011}.

\bibitem[{\citenamefont{Cook}(1994)}]{Cook94}
\bibinfo{author}{\bibfnamefont{G.~B.} \bibnamefont{Cook}},
  \bibinfo{journal}{Phys. Rev. D} \textbf{\bibinfo{volume}{50}},
  \bibinfo{pages}{5025} (\bibinfo{year}{1994}).

\bibitem[{\citenamefont{Baumgarte}(2000)}]{Baumgarte00a}
\bibinfo{author}{\bibfnamefont{T.~W.} \bibnamefont{Baumgarte}},
  \bibinfo{journal}{Phys. Rev. D} \textbf{\bibinfo{volume}{62}},
  \bibinfo{pages}{024018} (\bibinfo{year}{2000}), \eprint{gr-qc/0004050}.

\bibitem[{\citenamefont{Alcubierre et~al.}(2003)\citenamefont{Alcubierre,
  Br\"ugmann, Diener, Koppitz, Pollney, Seidel, and Takahashi}}]{Alcubierre02a}
\bibinfo{author}{\bibfnamefont{M.}~\bibnamefont{Alcubierre}},
  \bibinfo{author}{\bibfnamefont{B.}~\bibnamefont{Br\"ugmann}},
  \bibinfo{author}{\bibfnamefont{P.}~\bibnamefont{Diener}},
  \bibinfo{author}{\bibfnamefont{M.}~\bibnamefont{Koppitz}},
  \bibinfo{author}{\bibfnamefont{D.}~\bibnamefont{Pollney}},
  \bibinfo{author}{\bibfnamefont{E.}~\bibnamefont{Seidel}}, \bibnamefont{and}
  \bibinfo{author}{\bibfnamefont{R.}~\bibnamefont{Takahashi}},
  \bibinfo{journal}{Phys. Rev. D} \textbf{\bibinfo{volume}{67}},
  \bibinfo{pages}{084023} (\bibinfo{year}{2003}), \eprint{gr-qc/0206072}.

\bibitem[{\citenamefont{Alcubierre and Br\"ugmann}(2001)}]{Alcubierre00a}
\bibinfo{author}{\bibfnamefont{M.}~\bibnamefont{Alcubierre}} \bibnamefont{and}
  \bibinfo{author}{\bibfnamefont{B.}~\bibnamefont{Br\"ugmann}},
  \bibinfo{journal}{Phys. Rev. D} \textbf{\bibinfo{volume}{63}},
  \bibinfo{pages}{104006} (\bibinfo{year}{2001}), \eprint{gr-qc/0008067}.

\bibitem[{\citenamefont{Alcubierre
  et~al.}(2001{\natexlab{b}})\citenamefont{Alcubierre, Br\"ugmann, Pollney,
  Seidel, and Takahashi}}]{Alcubierre01a}
\bibinfo{author}{\bibfnamefont{M.}~\bibnamefont{Alcubierre}},
  \bibinfo{author}{\bibfnamefont{B.}~\bibnamefont{Br\"ugmann}},
  \bibinfo{author}{\bibfnamefont{D.}~\bibnamefont{Pollney}},
  \bibinfo{author}{\bibfnamefont{E.}~\bibnamefont{Seidel}}, \bibnamefont{and}
  \bibinfo{author}{\bibfnamefont{R.}~\bibnamefont{Takahashi}},
  \bibinfo{journal}{Phys. Rev. D} \textbf{\bibinfo{volume}{64}},
  \bibinfo{pages}{61501 (R)} (\bibinfo{year}{2001}{\natexlab{b}}),
  \eprint{gr-qc/0104020}.

\bibitem[{\citenamefont{Sperhake et~al.}(2003)\citenamefont{Sperhake, Smith,
  Kelly, Laguna, and Shoemaker}}]{Sperhake:2003fc}
\bibinfo{author}{\bibfnamefont{U.}~\bibnamefont{Sperhake}},
  \bibinfo{author}{\bibfnamefont{K.~L.} \bibnamefont{Smith}},
  \bibinfo{author}{\bibfnamefont{B.}~\bibnamefont{Kelly}},
  \bibinfo{author}{\bibfnamefont{P.}~\bibnamefont{Laguna}}, \bibnamefont{and}
  \bibinfo{author}{\bibfnamefont{D.}~\bibnamefont{Shoemaker}}
  (\bibinfo{year}{2003}), \eprint{gr-qc/0307015}.

\bibitem[{\citenamefont{Duez et~al.}(2003)\citenamefont{Duez, Marronetti,
  Shapiro, and Baumgarte}}]{Duez:2002bn}
\bibinfo{author}{\bibfnamefont{M.~D.} \bibnamefont{Duez}},
  \bibinfo{author}{\bibfnamefont{P.}~\bibnamefont{Marronetti}},
  \bibinfo{author}{\bibfnamefont{S.~L.} \bibnamefont{Shapiro}},
  \bibnamefont{and} \bibinfo{author}{\bibfnamefont{T.~W.}
  \bibnamefont{Baumgarte}}, \bibinfo{journal}{Phys. Rev.}
  \textbf{\bibinfo{volume}{D67}}, \bibinfo{pages}{024004}
  (\bibinfo{year}{2003}), \bibinfo{note}{gr-qc/0209102}.

\bibitem[{\citenamefont{Alcubierre
  et~al.}(2004{\natexlab{b}})\citenamefont{Alcubierre, Diener, Guzm\'an,
  Hawley, Koppitz, Pollney, and Seidel}}]{Alcubierre2003:co-rotating-shift}
\bibinfo{author}{\bibfnamefont{M.}~\bibnamefont{Alcubierre}},
  \bibinfo{author}{\bibfnamefont{P.}~\bibnamefont{Diener}},
  \bibinfo{author}{\bibfnamefont{F.~S.} \bibnamefont{Guzm\'an}},
  \bibinfo{author}{\bibfnamefont{S.}~\bibnamefont{Hawley}},
  \bibinfo{author}{\bibfnamefont{M.}~\bibnamefont{Koppitz}},
  \bibinfo{author}{\bibfnamefont{D.}~\bibnamefont{Pollney}}, \bibnamefont{and}
  \bibinfo{author}{\bibfnamefont{E.}~\bibnamefont{Seidel}}
  (\bibinfo{year}{2004}{\natexlab{b}}), \bibinfo{note}{in preparation}.

\bibitem[{\citenamefont{Br{\"u}gmann}()}]{BruegmannInPreparation}
\bibinfo{author}{\bibfnamefont{B.}~\bibnamefont{Br{\"u}gmann}},
  \bibinfo{note}{in preparation}.

\bibitem[{\citenamefont{Choptuik}(1993)}]{Choptuik93}
\bibinfo{author}{\bibfnamefont{M.~W.} \bibnamefont{Choptuik}},
  \bibinfo{journal}{Phys. Rev. Lett.} \textbf{\bibinfo{volume}{70}},
  \bibinfo{pages}{9} (\bibinfo{year}{1993}).

\bibitem[{\citenamefont{Plewa}()}]{AMRweb2}
\bibinfo{author}{\bibfnamefont{T.}~\bibnamefont{Plewa}},
  \emph{\bibinfo{title}{http://flash.uchicago.edu/$\sim$tomek/amr}}.

\bibitem[{\citenamefont{Diener et~al.}(2000)\citenamefont{Diener, Jansen,
  Khokhlov, and Novikov}}]{Diener99}
\bibinfo{author}{\bibfnamefont{P.}~\bibnamefont{Diener}},
  \bibinfo{author}{\bibfnamefont{N.}~\bibnamefont{Jansen}},
  \bibinfo{author}{\bibfnamefont{A.}~\bibnamefont{Khokhlov}}, \bibnamefont{and}
  \bibinfo{author}{\bibfnamefont{I.}~\bibnamefont{Novikov}},
  \bibinfo{journal}{Class. Quantum Grav.} \textbf{\bibinfo{volume}{17}},
  \bibinfo{pages}{435} (\bibinfo{year}{2000}), \bibinfo{note}{gr-qc/9905079}.

\bibitem[{\citenamefont{Choi et~al.}(2003)\citenamefont{Choi, Brown, Imbiriba,
  Centrella, and MacNeice}}]{Choi:2003ba}
\bibinfo{author}{\bibfnamefont{D.-I.} \bibnamefont{Choi}},
  \bibinfo{author}{\bibfnamefont{J.~D.} \bibnamefont{Brown}},
  \bibinfo{author}{\bibfnamefont{B.}~\bibnamefont{Imbiriba}},
  \bibinfo{author}{\bibfnamefont{J.}~\bibnamefont{Centrella}},
  \bibnamefont{and} \bibinfo{author}{\bibfnamefont{P.}~\bibnamefont{MacNeice}}
  (\bibinfo{year}{2003}), \eprint{physics/0307036}.

\bibitem[{\citenamefont{Choptuik et~al.}(2003)\citenamefont{Choptuik,
  Hirschmann, Liebling, and Pretorius}}]{Choptuik:2003ac}
\bibinfo{author}{\bibfnamefont{M.~W.} \bibnamefont{Choptuik}},
  \bibinfo{author}{\bibfnamefont{E.~W.} \bibnamefont{Hirschmann}},
  \bibinfo{author}{\bibfnamefont{S.~L.} \bibnamefont{Liebling}},
  \bibnamefont{and}
  \bibinfo{author}{\bibfnamefont{F.}~\bibnamefont{Pretorius}},
  \bibinfo{journal}{Phys. Rev.} \textbf{\bibinfo{volume}{D68}},
  \bibinfo{pages}{044007} (\bibinfo{year}{2003}), \eprint{gr-qc/0305003}.

\bibitem[{\citenamefont{Pretorius and Lehner}(2003)}]{Pretorius:2003wc}
\bibinfo{author}{\bibfnamefont{F.}~\bibnamefont{Pretorius}} \bibnamefont{and}
  \bibinfo{author}{\bibfnamefont{L.}~\bibnamefont{Lehner}}
  (\bibinfo{year}{2003}), \eprint{gr-qc/0302003}.

\bibitem[{\citenamefont{Br{\"u}gmann}(1996)}]{Bruegmann96}
\bibinfo{author}{\bibfnamefont{B.}~\bibnamefont{Br{\"u}gmann}},
  \bibinfo{journal}{Phys. Rev. D} \textbf{\bibinfo{volume}{54}},
  \bibinfo{pages}{7361} (\bibinfo{year}{1996}).

\bibitem[{\citenamefont{Lanfermann}(1999)}]{Lanfermann99}
\bibinfo{author}{\bibfnamefont{G.}~\bibnamefont{Lanfermann}}, Master's thesis,
  \bibinfo{school}{Freie Universit\"at Berlin, MPI f\"ur Gravitationsphysik}
  (\bibinfo{year}{1999}).

\bibitem[{\citenamefont{Schnetter et~al.}(2003)\citenamefont{Schnetter, Hawley,
  and Hawke}}]{Schnetter:2003rb}
\bibinfo{author}{\bibfnamefont{E.}~\bibnamefont{Schnetter}},
  \bibinfo{author}{\bibfnamefont{S.~H.} \bibnamefont{Hawley}},
  \bibnamefont{and} \bibinfo{author}{\bibfnamefont{I.}~\bibnamefont{Hawke}}
  (\bibinfo{year}{2003}), \eprint{gr-qc/0310042}.

\bibitem[{\citenamefont{Alcubierre et~al.}(2000)\citenamefont{Alcubierre,
  Brandt, Br{\"u}gmann, Gundlach, Mass{\'o}, Seidel, and
  Walker}}]{Alcubierre98b}
\bibinfo{author}{\bibfnamefont{M.}~\bibnamefont{Alcubierre}},
  \bibinfo{author}{\bibfnamefont{S.}~\bibnamefont{Brandt}},
  \bibinfo{author}{\bibfnamefont{B.}~\bibnamefont{Br{\"u}gmann}},
  \bibinfo{author}{\bibfnamefont{C.}~\bibnamefont{Gundlach}},
  \bibinfo{author}{\bibfnamefont{J.}~\bibnamefont{Mass{\'o}}},
  \bibinfo{author}{\bibfnamefont{E.}~\bibnamefont{Seidel}}, \bibnamefont{and}
  \bibinfo{author}{\bibfnamefont{P.}~\bibnamefont{Walker}},
  \bibinfo{journal}{Class. Quantum Grav.} \textbf{\bibinfo{volume}{17}},
  \bibinfo{pages}{2159} (\bibinfo{year}{2000}), \eprint{gr-qc/9809004}.

\bibitem[{\citenamefont{Cactus}()}]{Cactusweb}
\bibinfo{author}{\bibnamefont{Cactus}},
  \emph{\bibinfo{title}{http://www.cactuscode.org}}.

\end{thebibliography}

\end{document}